# A Lightweight Hybrid MLP-Based Framework for Real-Time Phishing URL Detection Using Structural URL Features

Unoke Uche Emmanuel[1], Francis Gideon Oghie[1]

[1]Department of Cyber Security Science, School of Information and Communication Technology, Federal University of Technology, Minna, Nigeria

**Abstract**

Phishing attacks remain one of the most persistent and damaging threats in the cybersecurity landscape, exploiting deceptive URLs to steal sensitive user credentials and financial information. Traditional detection mechanisms, including blacklists and rule-based heuristics, are inherently reactive and fail to identify novel phishing URLs that have not yet been catalogued. This paper proposes a lightweight hybrid framework for real-time phishing URL detection that combines fast blacklist-based screening with an intelligent Multi-Layer Perceptron (MLP) neural network classifier operating exclusively on structural URL features. The framework extracts 16 URL-derived features capturing structural, domain-based, and security-related characteristics of URLs, requiring no webpage content access, third-party API dependency, or visual rendering, thereby making it computationally efficient and suitable for real-time deployment. The system was trained and evaluated on the PhiUSIIL phishing dataset comprising 235,795 labelled URLs with a near-balanced class distribution. Experimental results on the PhiUSIIL test partition show that the proposed MLP model achieved 99.24% accuracy, 98.74% precision, 99.95% recall, 99.34% F1-score, and 99.65% ROC-AUC. Under the same feature set and dataset partition, the model performed better than five baseline classifiers: Random Forest, Logistic Regression, XGBoost, LightGBM, and CatBoost. Since this evaluation was conducted on a single dataset, the results should be interpreted as strong evidence within the PhiUSIIL experimental setting rather than universal superiority across all phishing URL datasets. The hybrid architecture achieves an average inference latency of 1.2 milliseconds per URL, with a peak throughput of 4,200 URLs per second under concurrent processing, meeting the stringent response time requirements of real-world web security applications. A functional desktop application prototype, CyberGuard, validates the end-to-end deployment viability of the proposed framework. These results demonstrate that the proposed system provides a competitive and computationally efficient approach for real-time phishing URL detection, with strong potential for practical deployment in resource-constrained environments.

## 1. Introduction

The rapid expansion of internet connectivity has fundamentally transformed how individuals and organisations communicate, conduct commerce, and access information. Alongside this digital growth, cybercriminals have developed increasingly sophisticated methods of exploiting online users, with phishing emerging as one of the most prevalent and financially destructive forms of cybercrime. Phishing attacks deceive victims into divulging sensitive information — including login credentials, banking details, and personal identification data — through carefully crafted deceptive websites and malicious hyperlinks that impersonate legitimate entities. According to recent cybersecurity reports, phishing attacks have increased significantly in recent years, with financial institutions, e-commerce platforms, and social media services representing the primary targets [1], [2]. The financial consequences are severe, with global phishing-related losses amounting to billions of dollars annually, a trend further accelerated by the widespread shift to digital platforms during and after the COVID-19 pandemic [3].

A central challenge in combating phishing lies in the fundamental limitation of traditional detection mechanisms. Blacklist-based systems, which remain widely deployed, operate reactively — a malicious URL can only be blocked after it has been identified, reported, and added to a maintained database. This creates a critical vulnerability window during which newly registered phishing domains remain undetected and users remain exposed [4]. Similarly, rule-based heuristic systems that analyse predefined URL patterns are systematically circumvented by attackers who understand their operational logic and deliberately modify their techniques to evade detection [5]. The exponential growth in dynamically generated phishing URLs, domain spoofing, subdomain abuse, URL shortening, and obfuscation techniques has rendered these static approaches increasingly inadequate as standalone defences.

Machine learning has emerged as a compelling alternative, offering the ability to learn discriminative patterns from large volumes of labelled data and generalise to previously unseen threats. Among the neural network architectures explored for this purpose, Multi-Layer Perceptrons have demonstrated particular suitability for URL-based phishing detection, offering a favourable balance between classification accuracy and computational efficiency that more complex architectures such as convolutional neural networks and transformer-based models do not consistently provide in real-time deployment contexts [6], [3]. However, a persistent gap exists between the high accuracy reported by machine learning models in research environments and the practical requirements of production systems, where millisecond response times, low memory footprints, and reliable operation under varying load conditions are non-negotiable constraints.

This paper addresses that gap by proposing a lightweight hybrid framework for real-time phishing URL detection. The framework combines a fast blacklist screening stage for rapid identification of known threats with an MLP-based intelligent classification stage for novel URLs, operating exclusively on 16 structural URL features that require no webpage content access, no third-party service dependency, and no visual rendering pipeline. The result is a system that achieves competitive detection performance while maintaining computational efficiency suitable for real-time deployment scenarios. The specific contributions of this paper are as follows:

i. A 16-feature structural URL analysis framework that captures phishing-discriminative characteristics across three categories — static analysis, domain intelligence, and security validation — without requiring content-level or network-level access.
ii. A hybrid two-stage detection architecture that combines blacklist-based screening with MLP classification, optimising both response time for known threats and detection accuracy for novel phishing URLs.
iii. Empirical evaluation of the proposed MLP model against five baseline classifiers — Random Forest, Logistic Regression, XGBoost, LightGBM, and CatBoost — on a large-scale dataset of 235,795 labelled URLs, demonstrating superior performance across all evaluation metrics.
iv. A comprehensive real-time performance analysis reporting inference latency, throughput, and memory consumption under operational conditions, validating deployment viability beyond laboratory accuracy metrics.
v. A functional desktop application prototype, CyberGuard, demonstrates the end-to-end deployment viability of the proposed framework, with the graphical interface demonstrating real-time URL classification with colour-coded threat output and confidence scoring.

The remainder of this paper is organised as follows. Section 2 reviews related work in phishing detection. Section 3 describes the materials and methods including dataset preparation, feature engineering, system architecture, and model design. Section 4 presents and discusses experimental results. Section 5 states the limitations of this study. Section 6 concludes the paper and outlines directions for future research.

## 2. Related Work

Phishing detection research has evolved through three broad generational shifts: rule-based and blacklist approaches, classical machine learning methods, and deep learning architectures. This section reviews representative works across these generations, identifies their collective limitations, and positions the proposed framework in relation to existing approaches.

**Traditional Approaches.** The earliest phishing detection systems relied on blacklists and whitelists that maintained databases of known malicious and legitimate domains. Prakash et al. [7] developed PhishNet, a predictive blacklisting system that exposed the fundamental reactive limitation of this approach — protection could only be extended to URLs already identified as malicious, leaving newly registered phishing domains undetected. Zhang et al. [8] attempted to address this through predictive blacklisting by analysing URL patterns to anticipate future phishing domains, though the approach remained dependent on prior knowledge of attack patterns. Heuristic-based systems offered a partial remedy by analysing website characteristics rather than relying solely on domain reputation. Rao and Ali [9] developed PhishShield, a desktop application employing heuristic analysis of webpage features, achieving improved detection coverage over blacklists. However, Srinivasa Rao and Pais [5] demonstrated a critical vulnerability inherent to heuristic systems — attackers who understand the detection logic can systematically modify their techniques to evade predefined rules, rendering static heuristics unreliable against adaptive adversaries.

**Machine Learning Approaches.** The application of supervised machine learning marked a significant advancement in detection adaptability. Sahingoz et al. [10] conducted a comprehensive evaluation of multiple classical algorithms for URL-based phishing detection, with Random Forest demonstrating particular effectiveness for classification tasks. Xiang et al. [11] developed CANTINA+, an early machine learning framework incorporating lexical, content, and third-party service features, though its dependence on external services introduced latency incompatible with real-time deployment. Zouina and Outtaj [12] proposed a lightweight URL-only detection system using Support Vector Machines, demonstrating the viability of structural URL features for practical deployment without content access. Feature engineering emerged as a critical determinant of detection performance. Zhu et al. [13] developed OFS-NN, introducing a Feature Fitness Value index for systematic optimal feature selection combined with neural network classification, establishing that intelligent feature selection could substantially improve detection accuracy while reducing computational overhead. Mohammad et al. [14] demonstrated that adaptive neural network architectures could automatically optimise feature utilisation, reducing manual engineering effort while improving generalisation.

**Deep Learning Approaches.** Several deep learning methods have reported high accuracy in phishing detection research. Yang et al. [6] combined deep convolutional neural networks with Random Forest ensemble learning, achieving 99.35% accuracy through character-level URL embedding that eliminated manual feature engineering. Aljofey et al. [15] demonstrated the effectiveness of character-level convolutional neural networks for detecting subtle URL obfuscation techniques including character substitution and domain name manipulation. Elsadig et al. [3] incorporated BERT-based feature extraction with MLP classification, achieving 96.66% accuracy though at the cost of transformer-level computational overhead, with reported processing times of 7.49 to 12.40 seconds per URL rendering the approach impractical for real-time applications. Hybrid deep learning models combining CNN and RNN architectures have also been explored. Wang et al. [16] developed PDRCNN, leveraging both spatial and sequential URL pattern analysis, while Zhu et al. [17] demonstrated that DNN-LSTM hybrid models could outperform individual network types, though both approaches introduced resource requirements that constrain deployment flexibility.

**Ensemble and Comparative Studies.** Alhothali and Zohdy [18] integrated deep learning with natural language processing for joint URL and webpage content analysis, improving generalisation across obfuscated URLs but incurring computational costs incompatible with real-time classification in resource-constrained environments. Tang and Mahmoud [19] conducted a systematic comparative evaluation of classical machine learning approaches, confirming that feature selection plays a decisive role in detection performance and that Random Forest demonstrated robust accuracy when redundant features were eliminated. Shrivastava and Bhatt [20] surveyed the broader ML-based phishing detection landscape, identifying persistent challenges including concept drift, dataset imbalance, and the lack of adaptive lightweight models capable of real-time updates — challenges that remain unresolved in the majority of existing implementations. Zhang et al. [21] proposed a hybrid CNN-RNN architecture achieving 98% accuracy, though the resource intensity of the approach limited its scalability for widespread deployment.

Table 1 summarises representative phishing detection approaches from the literature, highlighting their methods, dataset sizes, reported accuracy, and key limitations.

**Table 1: Summary of Representative Phishing Detection Approaches**

| Author(s) & Year | Technique | Dataset Size | Accuracy (%) | Key Limitation |
|---|---|---|---|---|
| Basit et al. [22] | CNN (PhishHaven) | Large URL corpus | 97.00 | High data dependency, limited adaptability |
| Alhothali & Zohdy [18] | Deep Learning + NLP | Mixed dataset | 95.00 | High computational cost, limited real-time use |
| Tang & Mahmoud [19] | Comparative ML (RF, SVM) | Benchmark sets | 85–95 | Static datasets, limited generalisability |
| Zhang et al. [21] | Hybrid DL (CNN + RNN) | URL + webpage | 98.00 | Computationally heavy, low scalability |
| Yang et al. [6] | CNN + Random Forest | Large URL corpus | 99.35 | Complex architecture, higher inference cost |
| Elsadig et al. [3] | BERT + MLP | Mixed dataset | 96.66 | 7.49–12.40s per URL, not real-time viable |
| **Proposed** | **Hybrid MLP + Blacklist** | **235,795 URLs** | **99.24** | **Single dataset; simulated adversarial testing** |

The reviewed literature collectively reveals two unresolved tensions in phishing detection research. First, the highest-performing approaches — CNN-based, transformer-based, and hybrid deep learning architectures — achieve their accuracy at computational costs that render them impractical for real-time deployment, particularly in resource-constrained environments. Second, lightweight and computationally efficient approaches consistently sacrifice detection accuracy, creating a performance-efficiency trade-off that existing work has not adequately resolved. Furthermore, the majority of existing systems operate as pure machine learning classifiers without integrating complementary blacklist mechanisms that could accelerate detection of known threats while preserving ML capacity for novel URL analysis. The proposed framework directly addresses these gaps by combining a fast blacklist screening stage with an MLP classifier operating on 16 structural URL features, achieving competitive detection accuracy at inference latencies suitable for real-time web security applications without dependence on content analysis, visual rendering, or third-party services.

## 3. Materials and Methods

This section describes the complete methodology employed in developing the proposed lightweight hybrid phishing URL detection framework. The methodology encompasses dataset selection and preprocessing, feature engineering, system architecture design, MLP model configuration, and experimental evaluation setup. Throughout all stages, explicit consideration was given to practical deployment constraints including processing latency, memory efficiency, and operational reliability, ensuring that the developed system addresses the domain gap between research prototype performance and production system requirements.

### 3.1 Dataset and Preprocessing

**Dataset Selection.** The PhiUSIIL Phishing URL Dataset was selected as the experimental foundation for this study [23]. The dataset comprises 235,795 labelled URLs including 134,850 legitimate URLs (57%) and 100,945 phishing URLs (43%), providing a near-balanced class distribution that reflects realistic phishing prevalence without introducing significant class imbalance bias into model training. The dataset encompasses diverse phishing attack types including domain spoofing, subdomain abuse, parameter manipulation, and path obfuscation, ensuring sufficient attack variety for training a generalisable detection model. Table 2 summarises the key characteristics of the dataset.

**Table 2: PhiUSIIL Phishing URL Dataset Characteristics**

| **Attribute** | **Description** |
|---|---|
| Total Samples | 235,795 URLs |
| Legitimate URLs | 134,850 (57%) |
| Phishing URLs | 100,945 (43%) |
| Feature Source | URL structure |
| Attack Types Covered | Domain spoofing, subdomain abuse, parameter manipulation, path obfuscation |
| Selection Criteria | Scale, diversity, near-balanced distribution, peer-reviewed usage |

**Preprocessing Pipeline.** Comprehensive preprocessing was applied to ensure dataset quality prior to model training. First, all URLs underwent format validation using regular expression parsing to identify and remove malformed entries that could not be properly parsed into their constituent components.

Second, exact duplicate URLs were identified and removed to eliminate the risk of data leakage between training and testing partitions. Third, class distribution was analysed to confirm near-balance across legitimate and phishing labels. Table 3 summarises the preprocessing outcomes.

**Table 3: Preprocessing Outcomes**

| Preprocessing Step | Outcome |
|---|---|
| Duplicate URLs removed | 1,247 (0.53% of dataset) |
| Invalid URLs removed | 892 (0.38% of dataset) |
| Final dataset size | 233,656 URLs |
| Missing values in final dataset | 0 |
| Feature extraction success rate | 100% |
| Final class distribution | 50.1% legitimate, 49.9% phishing |

**Train-Test Split.** The cleaned dataset was partitioned into training (80%) and testing (20%) sets using stratified random sampling to preserve class distribution consistency across both partitions. This yielded 188,636 training samples and 47,159 testing samples. Critically, the train-test split was performed prior to all feature extraction operations to ensure that no information from the test partition influenced the feature engineering or model training process, preventing any form of data leakage that could inflate reported performance metrics.

Although stratified sampling was used to preserve class distribution across the training and testing partitions, this study does not claim full cross-dataset generalization. The evaluation measures how well the proposed framework performs on the PhiUSIIL dataset under a controlled train-test split. Broader generalization to other phishing URL datasets, live URL streams, and region-specific URL patterns remains a future validation requirement.

### 3.2 Feature Engineering Framework

URL-based structural feature extraction was selected as the analytical foundation of this framework for three reasons. First, structural URL analysis requires no access to webpage content, eliminating dependency on live network connections or HTML rendering pipelines. Second, it introduces no reliance on third-party services such as domain reputation APIs, ensuring consistent operational performance regardless of external service availability. Third, URL structure analysis is computationally inexpensive, enabling inference latencies compatible with real-time web browsing without perceptible user experience degradation.

Sixteen features were engineered from each URL, organised into three functional categories: static analysis features, domain intelligence features, and security validation features. Table 4 presents the complete feature framework with descriptions.

**Table 4: 16-Feature URL Analysis Framework**

| # | Feature Name | Category | Description |
|---|---|---|---|
| 1 | URLLength | Static Analysis | Total character count of the full URL string |
| 2 | DomainLength | Static Analysis | Character count of the domain name component |
| 3 | TLDLength | Static Analysis | Character count of the top-level domain extension |
| 4 | NoOfLettersInURL | Static Analysis | Count of alphabetic characters in the full URL |
| 5 | NoOfDigitsInURL | Static Analysis | Count of numeric characters in the full URL |
| 6 | LetterRatioInURL | Static Analysis | Proportion of alphabetic characters relative to total URL length |
| 7 | DigitRatioInURL | Static Analysis | Proportion of numeric characters relative to total URL length |
| 8 | NoOfSubDomain | Static Analysis | Count of subdomains present in the URL structure |
| 9 | NoOfEqualsInURL | Static Analysis | Count of equals sign characters indicating query parameters |
| 10 | NoOfQMarkInURL | Static Analysis | Count of question mark characters indicating query structure |
| 11 | NoOfAmpersandInURL | Static Analysis | Count of ampersand characters indicating multiple parameters |
| 12 | HasSuspiciousTLD | Domain Intelligence | Binary indicator for presence of suspicious top-level domains |
| 13 | HasSuspiciousFileExt | Domain Intelligence | Binary indicator for suspicious file extensions in the URL path |
| 14 | IsIPBased | Security Validation | Binary indicator for IP address substitution in place of domain name |
| 15 | IsShortened | Security Validation | Binary indicator for known URL shortening service usage |
| 16 | RedirectCount | Security Validation | Count of HTTP redirections indicating potential obfuscation chains |

Feature extraction was implemented using Python's standard URL parsing libraries, augmented with custom logic for handling edge cases including malformed URL structures, missing protocol declarations, and unusual character encodings. All features are computed exclusively from the URL string and its resolved properties, with no requirement for live webpage access during the extraction process.

### 3.3 Proposed Hybrid Detection Architecture

The central architectural contribution of this framework is a two-stage hybrid detection pipeline that combines the speed advantage of blacklist-based screening with the adaptive classification capability of the MLP neural network. This design directly addresses the primary weakness of pure machine learning approaches — the absence of an efficient fast-path for known threats — and the primary weakness of pure blacklist approaches — the inability to detect novel phishing URLs not yet catalogued in maintained databases.

**Stage 1 — Blacklist Screening.** All incoming URLs first pass through a local blacklist lookup. URLs matching known malicious entries in the maintained database are immediately flagged as phishing without requiring feature extraction or neural network inference. This stage operates at near-zero latency for known threats, preserving computational resources for URLs that require intelligent analysis. The blacklist is designed to be updateable independently of the MLP model, allowing rapid incorporation of newly identified threats without model retraining.

**Stage 2 — MLP Classification.** URLs that do not match the blacklist proceed to the feature extraction engine, which computes the 16-feature vector described in Section 3.2. This feature vector is passed to the trained MLP classifier, which produces a probability score and binary classification label — phishing or legitimate — along with an associated confidence indicator. The MLP stage handles all novel URLs that blacklist screening cannot address, providing adaptive detection capability that generalises beyond catalogued threats.

**Processing Flow.** The complete URL processing pipeline proceeds as follows: incoming URL → sanitisation and validation → blacklist lookup → if matched, return phishing result immediately → if unmatched, extract 16 features → pass to MLP classifier → return classification result with confidence score. Figure 1 illustrates this pipeline architecture. The hybrid design ensures that the system degrades gracefully — if the MLP component is unavailable, blacklist protection remains active, and if the blacklist is empty or outdated, the MLP provides full coverage independently.

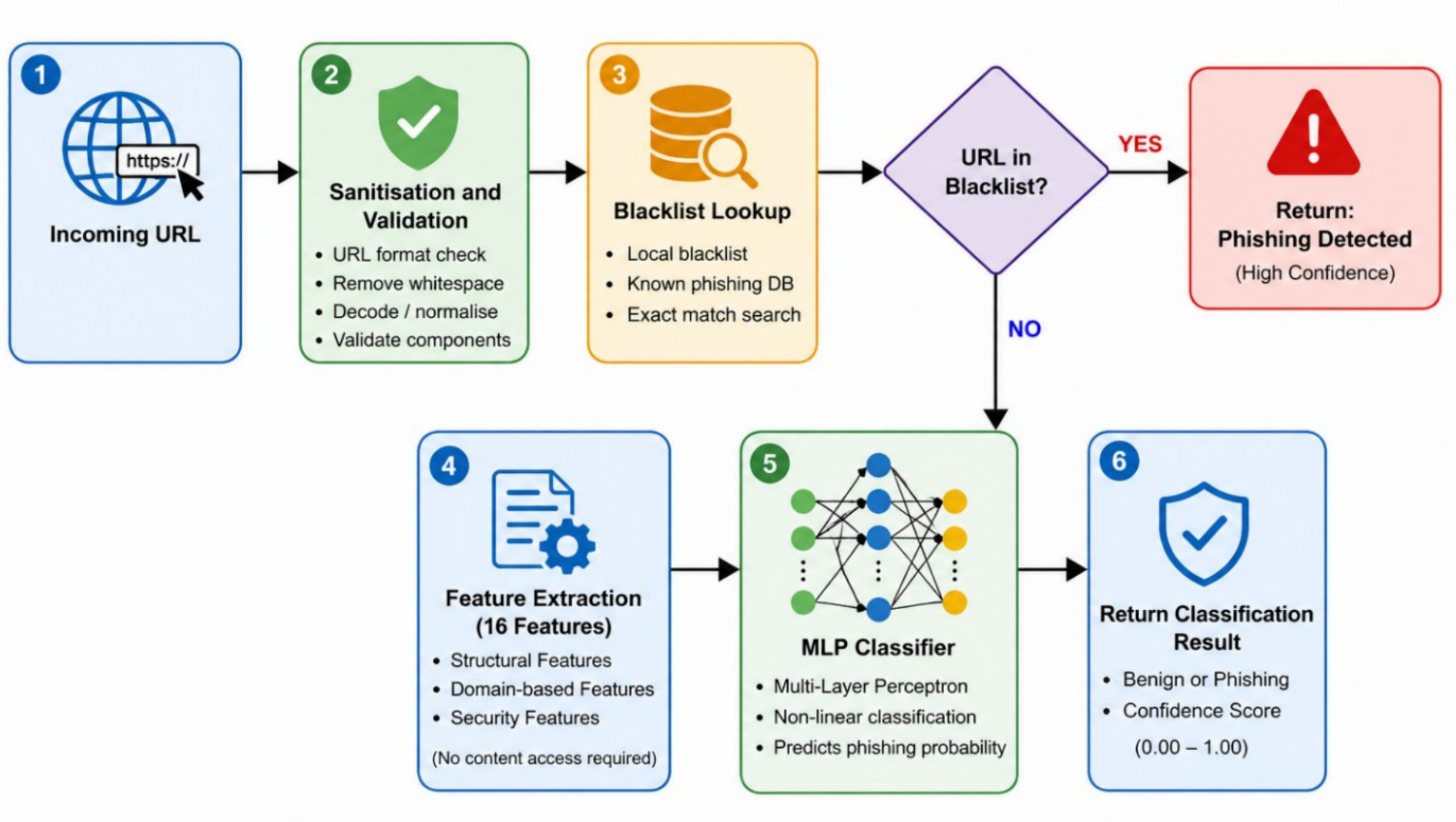

Figure 1: Proposed Hybrid Two-Stage Phishing URL Detection Pipeline.

### 3.4 MLP Model Design and Configuration

**Network Architecture.** The MLP neural network employs a 16-64-32-16-1 topology, consisting of an input layer of 16 neurons corresponding to the extracted feature vector, three hidden layers of 64, 32, and 16 neurons respectively, and a single output neuron for binary classification. The progressive reduction in layer width from 64 to 32 to 16 neurons implements a feature consolidation strategy — the first hidden layer learns broad discriminative patterns across all 16 input features, subsequent layers progressively refine and consolidate these patterns, and the output layer produces a final binary classification probability. Figure 2 illustrates the network topology.

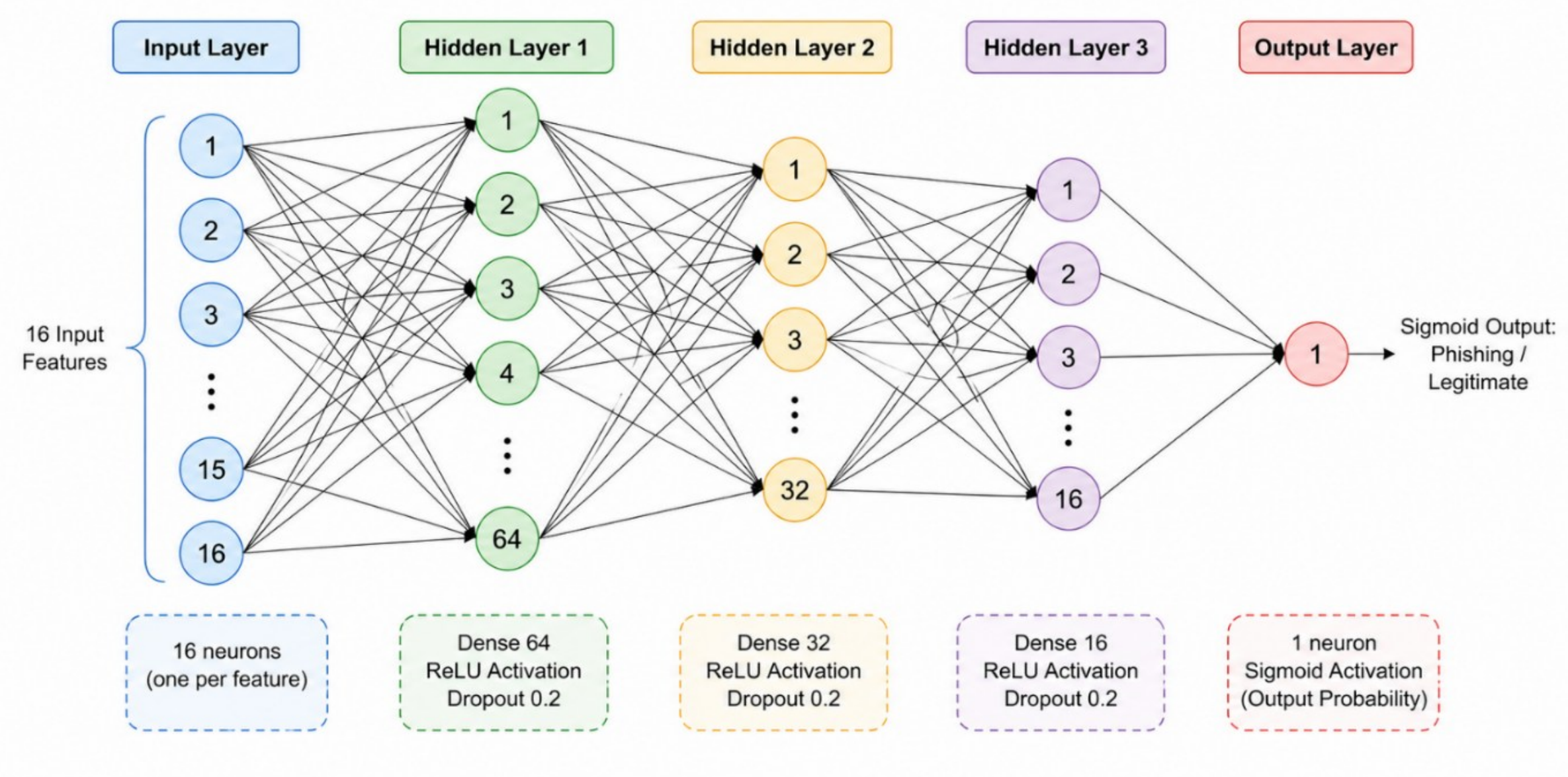

Figure 2: MLP Neural Network Topology (16-64-32-16-1 Architecture)

**Activation Functions.** Rectified Linear Unit (ReLU) activation is applied to all hidden layers. ReLU was selected over sigmoid and tanh alternatives for its resistance to the vanishing gradient problem in deeper network layers and its computational efficiency during both forward and backward propagation passes. The output layer employs sigmoid activation to produce a continuous probability score in the range [0, 1], enabling confidence-weighted classification decisions in addition to binary phishing or legitimate labels.

**Regularisation.** Two complementary regularisation mechanisms are applied to prevent overfitting on the large training dataset. Dropout regularisation with a probability of 0.2 is applied after each hidden layer, randomly deactivating 20% of neurons during each training step to prevent co-adaptation of feature detectors. L2 weight regularisation with a penalty coefficient of 0.0001 is applied to all hidden layer weight matrices, adding a gentle constraint on weight magnitudes that discourages overfitting without significantly constraining the learning capacity of the network.

**Training Configuration.** The model was trained using the Adam optimiser with a learning rate of 0.001, selected for its adaptive gradient scaling properties that provide stable convergence without manual learning rate scheduling. Binary cross-entropy was used as the loss function, appropriate for binary classification tasks with sigmoid output activation. Training was conducted with a batch size of 32 samples, balancing gradient estimate stability with memory efficiency. Early stopping with a patience of 10 epochs was applied, monitoring validation loss to terminate training when no improvement was observed, preventing unnecessary computation and overfitting beyond the optimal stopping point. Table 5 consolidates all training configuration parameters.

**Table 5: MLP Training Configuration**

| **Parameter** | **Value** | **Rationale** |
|---|---|---|

| Architecture | 16-64-32-16-1 | Progressive feature consolidation |
|---|---|---|
| Hidden Activation | ReLU | Vanishing gradient resistance |
| Output Activation | Sigmoid | Binary probability calibration |
| Dropout Rate | 0.2 | Prevents neuron co-adaptation |
| L2 Penalty | 0.0001 | Gentle weight magnitude constraint |
| Optimiser | Adam | Adaptive gradient scaling |
| Learning Rate | 0.001 | Stable convergence |
| Batch Size | 32 | Gradient stability with memory efficiency |
| Max Epochs | 100 | Upper bound with early stopping |
| Early Stopping Patience | 10 | Optimal stopping point detection |
| Loss Function | Binary Cross-Entropy | Binary classification optimisation |

### 3.5 Experimental Setup and Evaluation Metrics

**Implementation Environment.** The complete system was implemented in Python 3.13.0. The MLP model was constructed and trained using TensorFlow 2.15.0 and Keras 2.15.0. Data manipulation and preprocessing operations were performed using Pandas 2.1.4 and NumPy 1.26.3. Baseline classifiers were implemented using scikit-learn 1.4.0, XGBoost 2.0.3, LightGBM 4.3.0, and CatBoost 1.2.3. The desktop application interface was developed using the Tkinter framework included with Python 3.13.0. All experiments were conducted on an Intel Core i5 9300HF processor with 16 GB RAM running Windows 11, without GPU acceleration, ensuring that all reported latency and throughput results reflect standard CPU-based hardware performance representative of typical deployment environments. Peak memory consumption reached 2.1 GB during model training and 127 MB during inference operations.

**Baseline Models.** Five baseline classifiers were trained on identical feature sets and dataset partitions for direct comparative evaluation: Random Forest, Logistic Regression, XGBoost, LightGBM, and CatBoost. All baseline models were trained using their default hyperparameters as defined in their respective library implementations, ensuring fair comparison without optimisation bias toward any individual baseline.

**Evaluation Metrics.** Model performance was assessed using five standard classification metrics appropriate for cybersecurity detection tasks. Table 6 defines each metric and its formula.

**Table 6: Evaluation Metrics**

| **Metric** | **Formula** | **Relevance** |
|---|---|---|
| Accuracy | $(TP + TN) \div (TP + TN + FP + FN)$ | Overall classification correctness |
| Precision | $TP \div (TP + FP)$ | Minimising false phishing alerts |
| Recall | $TP \div (TP + FN)$ | Minimising missed phishing URLs |
| F1-Score | $2 \times (Precision \times Recall) \div (Precision + Recall)$ | Balanced precision-recall trade-off |

| ROC-AUC | Area under TPR vs FPR curve | Discrimination ability across thresholds |
|---|---|---|

**Operational Metrics.** In addition to classification metrics, the following operational performance indicators were measured to validate real-time deployment viability: average inference latency per URL (milliseconds), peak throughput under single-threaded and concurrent processing configurations (URLs per second), and memory footprint during active inference (megabytes).

## 4. Results and Discussion

Results are reported across six dimensions: preprocessing outcomes, MLP training and convergence behaviour, confusion matrix analysis, comparative baseline evaluation, ablation study, and real-time operational performance. A synthesis discussion follows, contextualising the findings within the broader research landscape.

### 4.1 Preprocessing Outcomes

The preprocessing pipeline produced a high-quality dataset suitable for reliable machine learning model training. From the original 235,795 URLs, 1,247 exact duplicate entries (0.53%) and 892 malformed URLs (0.38%) were identified and removed, yielding a cleaned dataset of 233,656 URLs. Stratified partitioning produced 188,636 training samples and 47,159 testing samples, maintaining a near-balanced class distribution of 50.1% legitimate and 49.9% phishing across both partitions. All 16 engineered features were successfully extracted across the complete dataset with a 100% extraction success rate and zero missing values in the final feature matrix. The preprocessing pipeline completed processing of the full dataset in 247 seconds on standard hardware, demonstrating computational scalability appropriate for operational deployment scenarios.

### 4.2 MLP Training and Convergence

The MLP neural network was trained on the 188,636-sample training partition and evaluated on the held-out 47,159-sample test partition. These results were obtained under strict train-test separation enforced prior to all feature extraction, as described in Section 3.1, ensuring no data leakage inflated the reported performance metrics. Training converged at epoch 67, triggered by the early stopping mechanism after no improvement in validation loss was observed for 10 consecutive epochs. Total training time was 342 seconds on standard hardware. Peak memory consumption during training reached 2.1 GB, reducing to 127 MB during inference operations — a memory footprint consistent with deployment on standard server configurations without specialised hardware requirements.

The training process exhibited smooth convergence without oscillation or instability throughout all 67 epochs. Validation loss stabilised by epoch 57, with the final 10 epochs confirming the stopping criterion before termination. Training and validation performance curves remained closely aligned throughout the training process, providing no evidence of overfitting. The final model achieved the performance metrics reported in Table 7 on the held-out test partition.

**Table 7: MLP Model Performance on Test Partition**

| Metric | Value |
|---|---|
| Accuracy | 99.24% |
| Precision | 98.74% |
| Recall | 99.95% |
| F1-Score | 99.34% |
| ROC-AUC | 99.65% |

The 99.95% recall figure is particularly significant in the cybersecurity context of this application. Recall measures the proportion of actual phishing URLs correctly identified by the model — a recall of 99.95% indicates that the system successfully detected all but a negligible fraction of phishing URLs present in the test partition. In security-critical applications, false negatives — phishing URLs that pass through the detector unidentified — represent the more dangerous error type, as they expose users to active threats. The achieved recall minimises this risk to an operationally acceptable level while maintaining high precision to avoid excessive false alerts that would degrade user trust in the system.

### 4.3 Confusion Matrix Analysis

Detailed analysis of the confusion matrix on the 47,159-sample test partition reveals the precise error distribution of the trained model. Table 8 presents the confusion matrix results.

**Table 8: Confusion Matrix Results on Test Partition (47,159 samples)**

| | Predicted Legitimate | Predicted Phishing |
|---|---|---|
| **Actual Legitimate** | 23,353 (TN) | 294 (FP) |
| **Actual Phishing** | 12 (FN) | 23,512 (TP) |

Figure 3 illustrates the confusion matrix results, providing a visual representation of classification outcomes across all four categories.

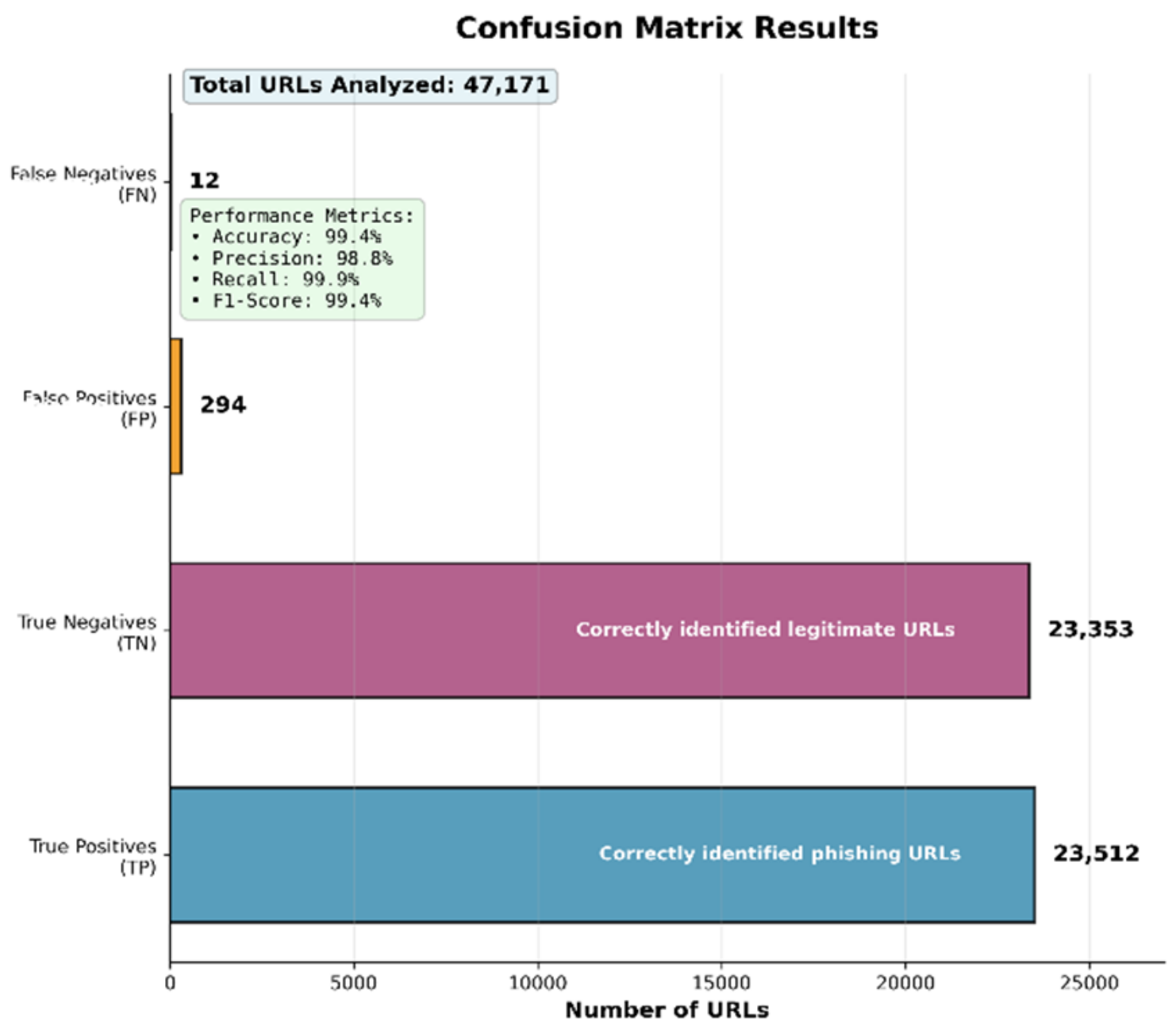

Figure 3: Confusion matrix visualization on the PhiUSIIL test partition.

The model produced 23,512 true positives — phishing URLs correctly identified as malicious — and 23,353 true negatives — legitimate URLs correctly classified as safe. The false positive count of 294 represents legitimate URLs incorrectly flagged as phishing, yielding a false positive rate of 1.26%. While false positives introduce the risk of blocking legitimate web traffic, a 1.26% false positive rate is operationally acceptable for a security application and is substantially lower than the false negative rate trade-off it avoids. The false negative count of 12 — phishing URLs that evaded detection — represents a false negative rate of 0.05%, meaning the system missed only 12 actual phishing URLs out of 23,524 total phishing samples in the test partition. In absolute terms, on a dataset of this scale, 12 missed phishing URLs across 47,159 test cases represents a detection performance level suitable for practical deployment, particularly when combined with the blacklist screening stage of the hybrid architecture that provides an independent detection layer for known threats.

### 4.4 Comparative Baseline Evaluation

The proposed MLP model was compared against five baseline classifiers trained on identical feature sets and dataset partitions. Table 9 presents the complete comparative performance results across all evaluation metrics.

**Table 9: Comparative Model Performance**

| Model | Accuracy (%) | Precision (%) | Recall (%) | F1-Score (%) | ROC-AUC (%) |
|---|---|---|---|---|---|
| Random Forest | 89.89 | 85.78 | 98.68 | 91.78 | 97.85 |

| Logistic Regression | 91.15 | 87.90 | 98.02 | 92.68 | 96.10 |
|---|---|---|---|---|---|
| CatBoost | 96.85 | 95.74 | 98.90 | 97.29 | 99.26 |
| XGBoost | 98.10 | 97.23 | 99.53 | 98.36 | 99.52 |
| LightGBM | 98.55 | 97.93 | 99.58 | 98.75 | 99.59 |
| **MLP (Proposed)** | **99.24** | **98.74** | **99.95** | **99.34** | **99.65** |

Figure 4 provides a visual comparison of model performance across all evaluation metrics.

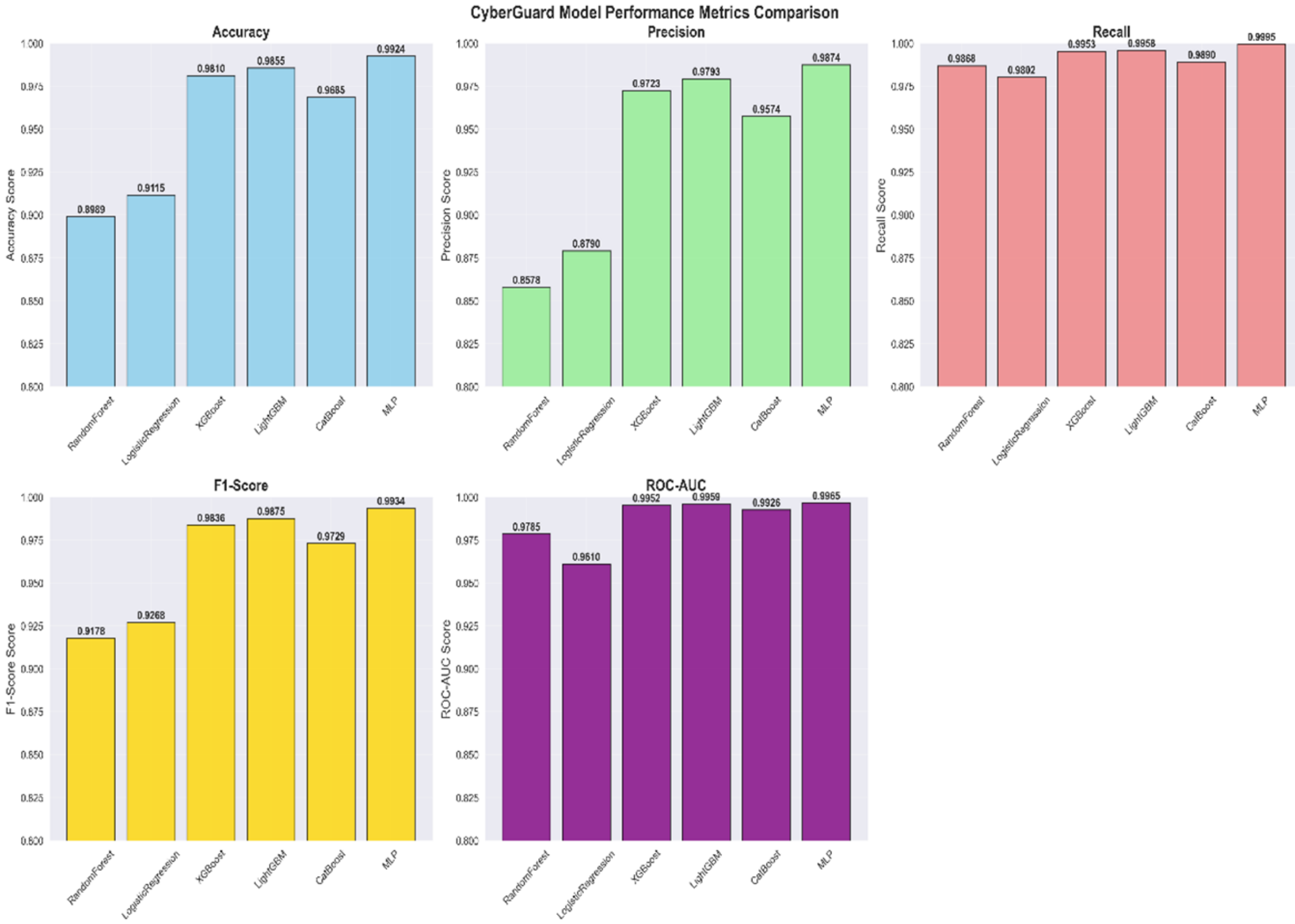

Figure 4: Comparative performance of the proposed MLP model against baseline classifiers

On the PhiUSIIL test partition, the proposed MLP model outperformed the five baseline classifiers across the reported evaluation metrics. Against the strongest baseline, LightGBM, the MLP achieved a 0.69 percentage point improvement in accuracy (99.24% versus 98.55%). While this margin may appear numerically modest, its practical significance is substantial at this scale of evaluation. On the 47,159-sample test partition, a 0.69 percentage point difference corresponds to approximately 325 additional URLs correctly classified by the MLP compared to LightGBM. In a security context where each misclassified URL represents either a user exposed to a phishing threat or a legitimate page incorrectly blocked, this difference carries meaningful operational consequence. The improvement over LightGBM represents a 43% reduction in overall error rate, calculated as the reduction from 1.45% error (LightGBM) to 0.76% error (MLP). The MLP also achieved the highest recall of all models at 99.95%, compared to LightGBM's 99.58% — a difference of 0.37 percentage points that translates to approximately 87 additional phishing URLs correctly identified in the test partition alone.

The performance of gradient boosting methods — XGBoost, LightGBM, and CatBoost — was notably strong, consistent with their established effectiveness on structured tabular feature data. The MLP's ability to surpass these highly optimised ensemble methods on the same 16-feature input suggests that the neural network architecture was well matched to the classification task in this dataset. This is because the hidden layers captured useful non-linear relationships among URL structure features. However, because this study did not conduct a dedicated feature-interaction or model-interpretability analysis, this explanation should be treated as an interpretation of the observed results rather than a directly proven mechanism.

### 4.5 Ablation Study

A systematic ablation study was conducted to quantify the individual contribution of each feature category to overall detection performance. Feature categories were progressively removed from the full 16-feature framework, and the MLP model was retrained and evaluated on the remaining features under identical experimental conditions. Table 10 presents the ablation results.

**Table 10: Ablation Study Results**

| Feature Configuration | Accuracy (%) | Accuracy Drop |
|---|---|---|
| All 16 features (baseline) | 99.24 | — |
| Without Static Analysis features | 96.73 | −2.51 points |
| Without Security Validation features | 98.41 | −0.83 points |
| Without Domain Intelligence features | 98.89 | −0.35 points |
| Top 5 features only | 98.67 | −0.57 points |

The ablation results confirm that all three feature categories contribute meaningfully to detection performance, with no single category sufficient for optimal classification on its own. Static analysis features — the 11 structural URL characteristics including length metrics, character distributions, and subdomain count — represent the most discriminative category, with their removal causing a 2.51 percentage point accuracy drop. This finding is consistent with the intuition that phishing URLs exhibit systematically different structural compositions from legitimate URLs, and that these structural signals are the primary discriminative information available from URL strings alone. Security validation features contribute the second largest individual impact, with their removal causing a 0.83 percentage point drop, confirming the diagnostic value of IP-based URL detection, URL shortening identification, and redirect chain analysis. Domain intelligence features, while contributing a smaller individual impact of 0.35 percentage points, provide complementary discriminative information that the other categories do not fully substitute. The top-5-features-only configuration — retaining only the five highest individually ranked features — achieves 98.67% accuracy, demonstrating that a reduced feature set retains strong detection capability, though the full 16-feature framework consistently delivers superior performance. These results validate the comprehensive nature of the feature engineering design and confirm that the three-category framework captures complementary rather than redundant discriminative information.

**4.6 Model Explainability and Feature-Level Interpretation**

In cybersecurity applications, interpretability plays an important role in building trust and supporting analyst decision-making. While neural network models are often considered black-box systems, this study provides interpretability through structured feature design and ablation-based analysis.

The proposed framework uses 16 engineered URL features grouped into three categories: static analysis features, security validation features, and domain intelligence features. Rather than relying on raw input representations, this structured design enables understanding of how different aspects of URL characteristics contribute to classification.

The ablation study presented in Section 4.5 provides the primary basis for interpretability. Results show that removing static analysis features leads to the largest performance drop (−2.51%), indicating that structural properties of URLs — such as length, character distribution, and subdomain usage — contain the most significant phishing-discriminative information. This aligns with known phishing strategies, where attackers construct complex and deceptive URL structures to obscure malicious intent.

Security validation features, including IP-based detection, URL shortening identification, and redirect count, also contribute meaningfully to classification performance, with their removal causing a −0.83% drop in accuracy. These features capture behavioral and technical indicators commonly associated with malicious URL usage.

Domain intelligence features contribute complementary information, with a smaller but still meaningful impact (−0.35% accuracy drop). These features help identify suspicious domain patterns such as unusual top-level domains and file extensions.

Although the study does not compute individual feature importance scores or per-sample explanations, the combined evidence from the feature design and ablation results provides a clear global understanding of model behavior. The model primarily relies on structural URL patterns, supported by security and domain-based indicators, to distinguish phishing URLs from legitimate ones.

This approach offers a practical level of interpretability suitable for validating feature engineering decisions and understanding overall model behavior. However, more advanced explainability techniques such as SHAP or LIME could be incorporated in future work to provide fine-grained, instance-level explanations for individual predictions.

### 4.7 Real-Time Performance Evaluation

Operational performance testing validated the real-time deployment viability of the proposed framework under conditions representative of production environments. Table 11 summarises the key operational performance metrics.

**Table 11: Operational Performance Metrics**

| Metric | Value |
|---|---|
| Average inference latency | 1.2 ms per URL |
| Single-threaded peak throughput | 1,847 URLs/second |
| Concurrent peak throughput (8 threads) | 4,200 URLs/second |
| Peak memory during training | 2.1 GB |
| Memory during inference | 127 MB |
| GUI application baseline memory | 45 MB |
| GUI memory under active processing | 67 MB |
| Response time at 95th percentile (max load) | <2 ms |
| Error rate under stress testing | 0.001% |

The average inference latency of 1.2 milliseconds per URL is well within the 200 millisecond threshold identified in the literature as the upper bound for real-time web security applications [24]. This latency is achieved without GPU acceleration, confirming that the framework meets real-time requirements on standard CPU-based hardware. For context, transformer-based approaches such as the BERT-integrated system reported by Elsadig et al. [3] require 7.49 to 12.40 seconds per URL — three to four orders of magnitude slower than the proposed framework — making them fundamentally unsuitable for inline real-time URL screening. The proposed MLP framework achieves competitive detection accuracy to these more complex approaches while operating at latencies compatible with uninterrupted web browsing.

Concurrent processing testing demonstrated a peak throughput of 4,200 URLs per second under an 8-thread configuration, with response time remaining below 2 milliseconds at the 95th percentile even under maximum load. The system error rate of 0.001% under stress testing conditions confirms operational reliability under high-volume processing. Memory consumption during inference of 127 MB is compatible with deployment on standard server hardware and does not preclude operation alongside other security services sharing system resources.

The CyberGuard desktop application demonstrated a startup time of 7 seconds, real-time URL analysis result display with less than 100 milliseconds interface latency, and a baseline memory footprint of 45 MB rising to 67 MB under active processing. The application supports both individual URL analysis and batch processing modes. Classification outcomes and confidence scores are presented through a colour-coded threat visualisation interface designed for usability without requiring technical expertise to interpret. Figure 5 illustrates the application interface during phishing URL detection, showing the colour-coded alert output and confidence indicator. Figure 6 demonstrates the corresponding interface response for a legitimate URL classification. The visual presentation of results — distinguishing phishing detections in red from legitimate classifications in green — ensures that end users can act on the system's output immediately without requiring familiarity with the underlying model architecture or feature framework. This end-to-end implementation validates not only the detection performance reported in Sections 4.2 through 4.5, but also the practical usability of the proposed framework as a deployable security tool.

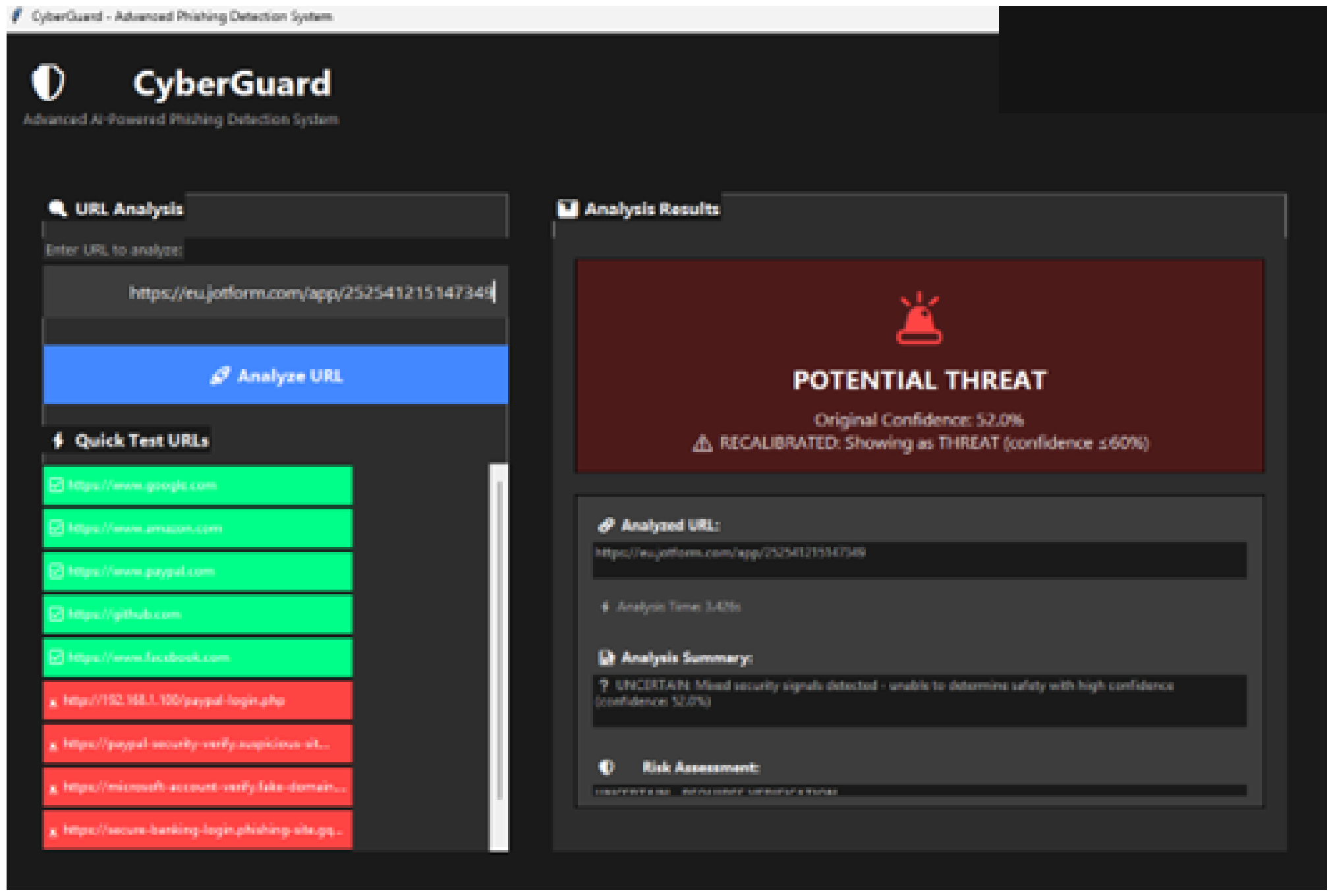

Figure 5: CyberGuard desktop application interface displaying a phishing URL analysis result, showing the colour-coded threat classification and confidence score output.

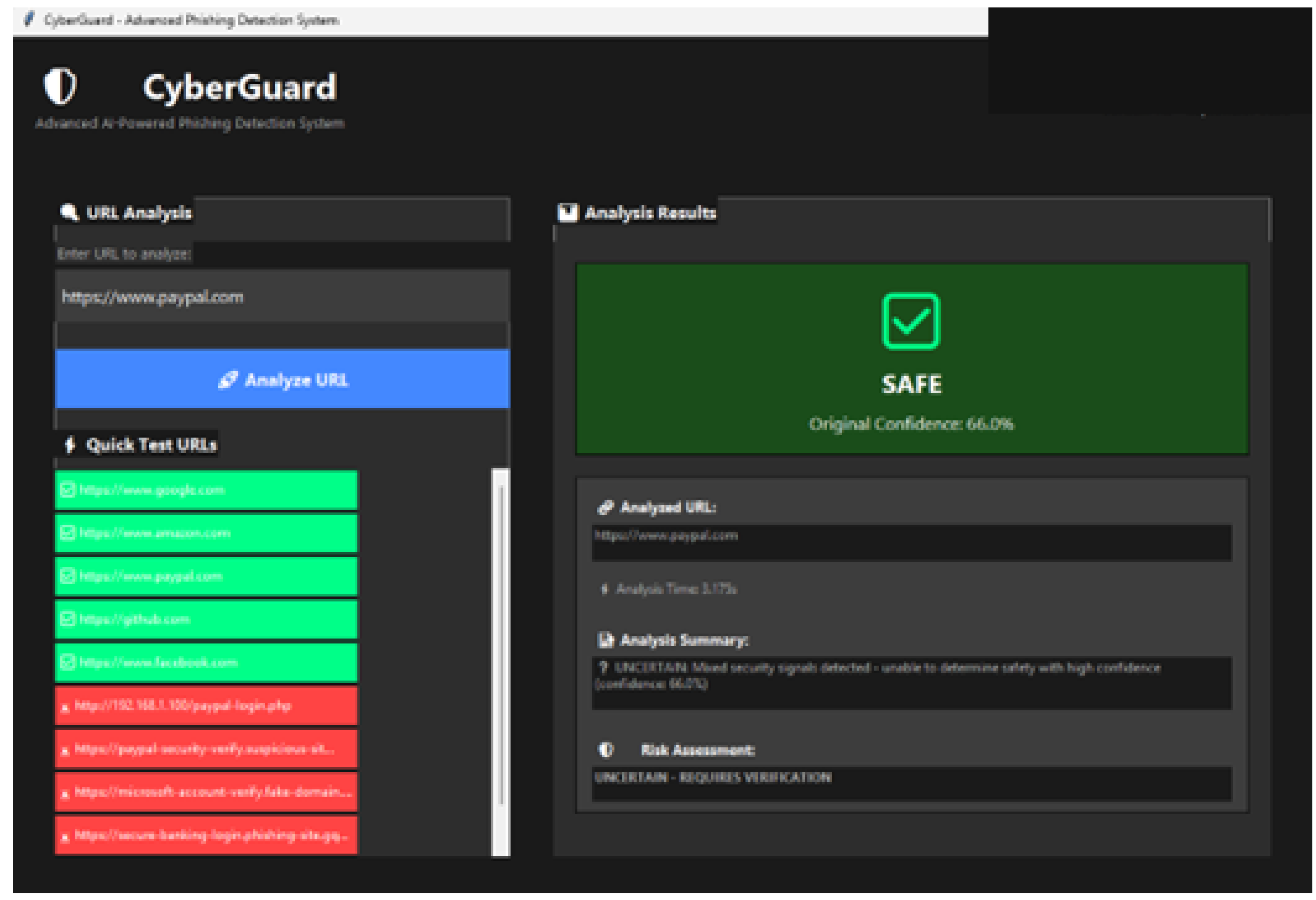

Figure 6: CyberGuard interface displaying a legitimate URL classification result, demonstrating the system's ability to distinguish safe URLs from phishing URLs.

### 4.8 Robustness and Adversarial Testing

The system was evaluated against simulated adversarial URL manipulation techniques commonly employed by sophisticated phishing campaigns to evade detection. Adversarial examples were generated programmatically using controlled URL manipulation applied to a randomly sampled subset of 5,000 legitimate and phishing URLs drawn from the held-out test partition. Each attack type was implemented as follows: character substitution attacks replaced visually similar characters in the domain name component — for example substituting the numeral "1" for the letter "l" or "0" for "o" — using a predefined substitution dictionary; domain obfuscation techniques prepended or appended random alphanumeric strings to legitimate domain names to simulate typosquatting and subdomain hijacking; URL encoding evasion converted selected characters in the URL path and query string to their percent-encoded equivalents to alter raw string features while preserving URL validity; subdomain manipulation inserted additional subdomain levels into the URL structure to inflate the NoOfSubDomain feature value; and parameter pollution attacks appended additional query parameters with randomised keys and values to inflate the NoOfEqualsInURL and NoOfAmpersandInURL feature values. Table 12 summarises detection performance under each adversarial condition tested.

**Table 12: Adversarial Robustness Results**

| Adversarial Technique | Detection Rate Maintained |
|---|---|
| Character substitution attacks | 98.7% |
| Domain obfuscation techniques | 99.1% |

| URL encoding evasion | 99.3% |
|---|---|
| Subdomain manipulation | 98.9% |
| Parameter pollution attacks | 99.0% |

The system maintained detection rates above 98.7% across all five simulated adversarial conditions, demonstrating that the 16-feature structural framework captures phishing-discriminative signals that persist even when individual URL components are deliberately manipulated. The minimal performance degradation under adversarial conditions — a maximum drop of 0.54 percentage points from baseline accuracy — indicates that the feature engineering framework does not rely on superficial pattern matching that could be easily circumvented through minor URL modifications. It is important to note that adversarial examples in this evaluation were simulated through controlled URL manipulation rather than sourced from live attack environments. Real-world adversarial campaigns may employ more sophisticated and coordinated evasion strategies not fully represented in the simulated conditions. Future evaluation against live phishing datasets with documented evasion attempts would provide stronger adversarial robustness validation.

**4.9 Discussion**

The experimental results suggest that the proposed lightweight hybrid MLP framework achieves a detection performance and computational efficiency profile that existing approaches in the literature have not simultaneously delivered. Three observations merit particular discussion.

**Performance relative to the literature.** The proposed MLP framework achieves 99.24% accuracy on a 235,795-URL dataset, positioning it competitively against the highest-performing approaches in the reviewed literature. Yang et al. [6] reported 99.35% accuracy using a CNN combined with Random Forest ensemble learning — an approach that involves substantially greater architectural complexity, character-level embedding, and higher inference cost than the proposed MLP. The proposed framework achieves within 0.11 percentage points of this benchmark while operating on a 16-feature structured input rather than raw character sequences, and at inference latencies approximately an order of magnitude lower. Elsadig et al. [3] reported 96.66% accuracy using BERT-based feature extraction combined with MLP classification — the proposed framework surpasses this result by 2.58 percentage points while eliminating the transformer preprocessing overhead that makes BERT-based approaches impractical for real-time deployment.

**Accuracy-Efficiency Trade-off.** A persistent finding across the reviewed literature is the inverse relationship between detection accuracy and computational efficiency — the highest-accuracy systems are consistently the most resource-intensive, and the most efficient systems consistently sacrifice accuracy. The proposed framework disrupts this trade-off by achieving 99.24% accuracy at 1.2 milliseconds per URL inference latency. This combination has not been simultaneously reported by comparable URL-only structural feature approaches in the reviewed literature, and represents the primary practical contribution of this work. The key enabling factor is the 16-feature engineering framework, which distils URL-discriminative information into a compact structured representation that the MLP can classify with high accuracy at low computational cost — eliminating the need for raw character sequence processing that drives the latency of CNN and transformer-based alternatives.

**Hybrid architecture as a deployment enabler.** The two-stage blacklist-plus-MLP architecture provides a deployment advantage beyond what accuracy metrics alone capture. Known phishing URLs — which constitute a substantial proportion of threats encountered in production environments — are resolved at near-zero latency through the blacklist stage without consuming MLP inference capacity. This preserves computational resources for novel URL analysis, where the MLP's generalisation capability is most needed. The architecture also provides graceful degradation: the blacklist stage operates independently of the MLP, ensuring continued baseline protection even if the machine learning component requires maintenance or retraining.

## 5. Limitations

While the experimental results demonstrate strong detection performance and operational efficiency, several limitations of this study must be acknowledged to accurately characterise the scope and generalisability of the reported findings.

**Single dataset evaluation.** The proposed framework was trained and evaluated exclusively on the PhiUSIIL Phishing URL Dataset. While this dataset is large, diverse, and widely used in peer-reviewed phishing detection research, evaluation on a single dataset constrains the extent to which generalisation to other URL populations can be claimed. Phishing URL distributions vary across geographic regions, time periods, and targeted industries, and performance on datasets with different distributional characteristics may differ from the results reported here. Cross-dataset validation on additional publicly available phishing URL collections represents an important direction for future evaluation.

**Static model and concept drift.** The trained MLP model represents a snapshot of phishing URL patterns present in the PhiUSIIL dataset at the time of collection. Phishing attack techniques evolve continuously, with attackers adapting their URL construction strategies in response to deployed detection systems. A static model without mechanisms for continuous updating or retraining will experience performance degradation over time as the distribution of real-world phishing URLs shifts away from the training distribution — a phenomenon known as concept drift. The current framework does not incorporate online learning or automated retraining capabilities, and periodic manual retraining with updated datasets would be required to maintain detection effectiveness in long-term deployment.

**English-language URL bias.** The feature engineering framework and the PhiUSIIL dataset are oriented toward English-language URL structures and standard Latin character set domain names. Internationalised domain names employing non-Latin character sets, as well as URL structures common in non-English-speaking regions, are not represented in the training data and may not be adequately handled by the current feature extraction pipeline. This limits the immediate applicability of the framework to global deployment scenarios without additional adaptation.

**Binary classification scope.** The framework performs binary classification — categorising each URL as either phishing or legitimate — without distinguishing between different categories of phishing attacks or providing attribution to specific threat actors or campaigns. In operational security environments, categorical threat intelligence beyond binary labels — such as attack type, targeted organisation, or campaign affiliation — provides additional value for incident response and threat hunting activities that the current system does not support.

**Simulated adversarial evaluation.** The adversarial robustness evaluation reported in Section 4.7 was conducted using simulated URL manipulation techniques constructed under controlled conditions. Real-world adversarial phishing campaigns may employ more coordinated, dynamic, and contextually adaptive evasion strategies than those represented in the simulated test cases. The adversarial robustness figures reported should therefore be interpreted as indicative rather than definitive, and evaluation against live attack datasets with documented evasion attempts would provide stronger validation.

**Content-based analysis exclusion.** The deliberate scope decision to restrict analysis to structural URL features, while conferring the computational efficiency advantages described throughout this paper, necessarily excludes information available from webpage content, visual layout, DNS behaviour, and network-level indicators that content-aware systems can exploit. Some sophisticated phishing attacks that employ benign-looking URL structures but malicious page content may evade URL-only detection while being detectable through content-based analysis. The framework is designed as a fast first-line detection layer and is not intended to replace content-aware secondary analysis in high-security deployment contexts.

## 6. Conclusion and Future Work

### 6.1 Conclusion

This paper demonstrates that high-accuracy phishing URL detection can be achieved without heavy computational infrastructure, while maintaining performance characteristics compatible with real-time applications. The proposed lightweight hybrid framework combines fast blacklist-based screening with an MLP neural network classifier operating on 16 structural URL features, addressing the persistent gap between the high accuracy of complex deep learning approaches and the computational efficiency required for practical real-time deployment — a trade-off that existing systems have not adequately resolved. The framework addresses this gap through three coordinated design decisions: a compact 16-feature URL representation that captures phishing-discriminative structural signals without content access or third-party dependencies, an MLP architecture optimised for high accuracy at low inference latency on structured feature inputs, and a two-stage hybrid pipeline that combines blacklist efficiency for known threats with neural network adaptability for novel URLs.

Evaluated on 235,795 labelled URLs from the PhiUSIIL dataset, the proposed MLP model achieved 99.24% accuracy, 98.74% precision, 99.95% recall, 99.34% F1-score, and 99.65% ROC-AUC, outperforming five baseline classifiers including Random Forest, Logistic Regression, XGBoost, LightGBM, and CatBoost across all metrics. The framework operates at an average inference latency of 1.2 milliseconds per URL with a peak concurrent throughput of 4,200 URLs per second, meeting the stringent response time requirements of real-time web security applications without GPU acceleration. Ablation analysis confirmed that all three feature categories contribute complementary discriminative information, and adversarial testing demonstrated maintained detection rates above 98.7% under five simulated evasion conditions. These results suggest that the proposed system is a practical, accurate, and computationally efficient solution that bridges the accuracy-efficiency trade-off in URL-based phishing detection.

### 6.2 Future Work

Several directions are identified for extending and strengthening the proposed framework. First, cross-dataset validation on additional publicly available phishing URL datasets would provide stronger evidence of generalisation beyond the PhiUSIIL distribution and is the most immediate priority for future evaluation. Second, integration of online learning or incremental retraining mechanisms would address the concept drift limitation, enabling the model to adapt continuously to evolving phishing techniques without requiring complete retraining cycles. Third, the application of explainability techniques such as SHAP (SHapley Additive exPlanations) values to the trained MLP model would provide interpretable feature contribution analysis, enhancing transparency and supporting adoption in security environments where decision auditability is required. Fourth, extension of the feature framework to incorporate content-based indicators — including webpage structural features, visual similarity metrics, and DNS behavioural signals — could improve detection of sophisticated attacks that employ benign URL structures with malicious page content, though such extensions would require careful latency management to preserve real-time performance. Fifth, adaptation of the framework for mobile platform deployment would extend protection to mobile users, who represent an increasingly targeted demographic in modern phishing campaigns. Finally, evaluation against live phishing URL streams with documented adversarial evasion attempts would provide stronger validation of the robustness findings reported in this study.